\documentclass[prb,twocolumn,showpacs,superscriptaddress]{revtex4}

\usepackage{graphicx}
\usepackage{dcolumn}
\usepackage{amsmath}
\usepackage{color}

\newcommand{\SrCrO}{Sr$_{3}$Cr$_2$O$_8$}
\newcommand{\BaCrO}{Ba$_{3}$Cr$_2$O$_8$}
\newcommand{\ACrO}{$A_{3}$Cr$_2$O$_8$}

\begin{document}

\title{High-field spectroscopy of singlet-triplet transitions in the spin-dimer systems Sr$_{3}$Cr$_2$O$_8$ and Ba$_{3}$Cr$_2$O$_8$}

\author{Zhe~Wang}
\affiliation{Experimental Physics V, Center for Electronic
Correlations and Magnetism, Institute of Physics, University of Augsburg, 86135 Augsburg, Germany}

\author{D. Kamenskyi}
\altaffiliation[Present address: ]
{Radboud University Nijmegen, Institute for Molecules and Materials, High Field Magnet Laboratory, 6500 GL Nijmegen, The Netherlands.}
\affiliation{Dresden High Magnetic Field Laboratory (HLD), Helmholtz-Zentrum
Dresden-Rossendorf, 01328 Dresden, Germany}

\author{O. C\'{e}pas}
\affiliation{Institut N\'{e}el, CNRS and Universit\'{e} Joseph Fourier, BP 166, 38042 Grenoble Cedex 9, France}

\author{M.~Schmidt}
\affiliation{Experimental Physics V, Center for Electronic
Correlations and Magnetism, Institute of Physics, University of Augsburg, 86135 Augsburg, Germany}

\author{D. L. Quintero-Castro}
\affiliation{Helmholtz-Zentrum Berlin f\"{u}r Materialien und
Energie, 14109 Berlin, Germany}

\author{A. T. M. N. Islam}
\affiliation{Helmholtz-Zentrum Berlin f\"{u}r Materialien und
Energie, 14109 Berlin, Germany}

\author{B. Lake}
\affiliation{Helmholtz-Zentrum Berlin f\"{u}r Materialien und
Energie, 14109 Berlin, Germany}
\affiliation{Institut f\"{u}r Festk\"{o}rperphysik, Technische
Universit\"{a}t Berlin, 10623 Berlin, Germany}

\author{A. A. Aczel}
\affiliation{Department of Physics and Astronomy, McMaster University, Hamilton, Ontario L8S 4M1, Canada}
\affiliation{Quantum Condensed Matter Division, Oak Ridge National Laboratory, Oak Ridge, TN 37831, USA}
\author{H. A. Dabkowska}
\author{A. B. Dabkowski}
\affiliation{Brockhouse Institute for Materials Research, McMaster University, Hamilton, Ontario L8S 4M1, Canada}

\author{G. M. Luke}
\affiliation{Department of Physics and Astronomy, McMaster University, Hamilton, Ontario L8S 4M1, Canada}

\author{Yuan~Wan}
\affiliation{Department of Physics and Astronomy, Johns Hopkins University, Baltimore, Maryland 21218, USA}

\author{A.~Loidl}
\affiliation{Experimental Physics V, Center for Electronic
Correlations and Magnetism, Institute of Physics, University of Augsburg, 86135 Augsburg, Germany}

\author{M. Ozerov}
\affiliation{Dresden High Magnetic Field Laboratory (HLD), Helmholtz-Zentrum
Dresden-Rossendorf, 01328 Dresden, Germany}
\author{J. Wosnitza}
\affiliation{Dresden High Magnetic Field Laboratory (HLD), Helmholtz-Zentrum
Dresden-Rossendorf, 01328 Dresden, Germany}
\affiliation{Institut f\"{u}r Festk\"{o}rperphysik, Technische
Universit\"{a}t Dresden, 01068 Dresden, Germany}

\author{S. A. Zvyagin}
\affiliation{Dresden High Magnetic Field Laboratory (HLD), Helmholtz-Zentrum
Dresden-Rossendorf, 01328 Dresden, Germany}

\author{J.~Deisenhofer}
\affiliation{Experimental Physics V, Center for Electronic
Correlations and Magnetism, Institute of Physics, University of Augsburg, 86135 Augsburg, Germany}

\date{\today}

\begin{abstract}
Magnetic excitations in the isostructural spin-dimer systems \SrCrO~and \BaCrO~are probed by means of high-field electron spin resonance at sub-terahertz frequencies.  Three types of magnetic modes  were observed.  One mode  is gapless and  corresponds to transitions within excited states, while two other  sets  of modes  are gapped and  correspond to transitions from the ground to the first excited states.  The selection rules of the gapped modes  are analyzed  in terms of a dynamical Dzyaloshinskii-Moriya interaction, suggesting the presence of phonon-assisted effects in the low-temperature spin dynamics of \SrCrO~and \BaCrO.

\end{abstract}


\maketitle

\section{Introduction}
High-field electron spin resonance (ESR) is a very
powerful mean to study the excitation spectrum
and the transition matrix elements
 resulting from the coupling between
radiation and matter.  In magnetic systems, the transitions do not
always result from the Zeeman coupling of the spins to the
magnetic field of the radiation but may result from indirect
processes,\cite{ElliottL,Shen} thus providing information on these couplings.

Magnetic systems consisting of a small number of interacting magnetic
moments in a cluster are particularly interesting in this respect.
The simplest cluster is a spin dimer of two spin-$1/2$ ions
coupled by an antiferromagnetic Heisenberg interaction $J_0>0$, leading to a singlet
ground state ($S=0$) separated from a triplet excitation ($S=1$) by an
energy gap.  In general, because the dimers are regularly arranged in
a crystal and coupled to their neighbours, the triplet excitations acquire a
dispersion. However, the overall simple picture of the excitation spectrum
may remain the same if the interactions are weak or frustrated.

Singlet-triplet transitions have been observed by inelastic neutron
scattering and high-field ESR
measurements in many spin-dimer antiferromagnets, e.g.,
SrCu$_2$(BO$_3$)$_2$\cite{Nojiri99,Room04} and
CuTe$_2$O$_5$\cite{Wang11b} based on Cu$^{2+}$ ($3d^9$, $s=1/2$) ions,
and \BaCrO\cite{Kofu09a,Kofu09b,Kamenskyi13} and
\SrCrO\cite{Castro10,Wang11,Castro12} based on Cr$^{5+}$ ($3d^1$, $s=1/2$)
ions. While the dispersion as a function of wavevector and energy can be measured, for example, by single crystal inelastic neutron scattering, photons of the relevant energy usually probe only the $\Gamma$ point.  Moreover, such
direct singlet-triplet transitions with $\Delta S=1$ (magnon-like) are
optically forbidden. Their observation
implies that the total spin $S$ is not a good quantum number, i.e., rotation
symmetry in spin-space is broken. This naturally arises when
spin-orbit coupling is present, but the effects are weak for
$3d$ transition metal ions.

\begin{figure}[t]
\centering
\includegraphics[width=85mm,clip]{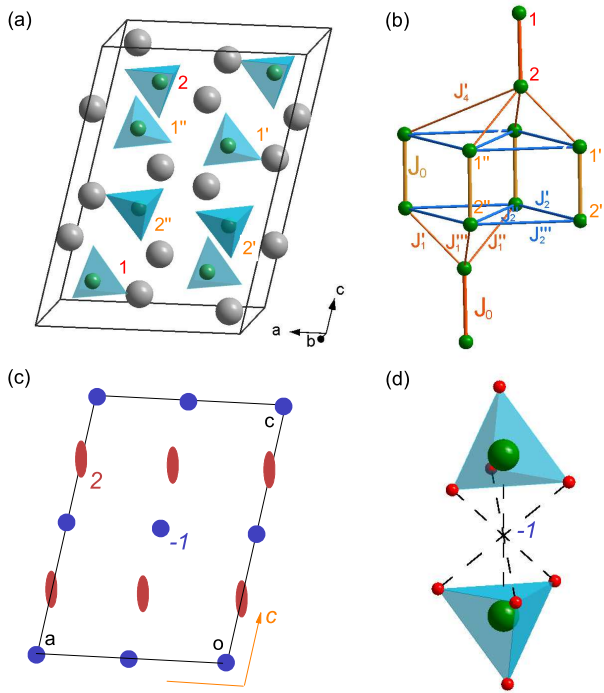}
\vspace{2mm} \caption[]{\label{Fig:STScheme} (Color online) (a) Unit cell of \ACrO~($A$ = Sr,Ba) in the low-temperature monoclinic phase with space group $C2/c$. (b) Layered structure of spin dimers with anisotropic exchange interactions from Ref.~\onlinecite{Castro10}. $J_0$ is the intra-dimer exchange interaction. Cr pairs are labeled by $(12)$, $(1'2')$ and $(1''2'')$ in accord with (a).  (c) Symmetry elements of the space group $C2/c$: inversion center $-$\emph{1}, two-fold rotation \emph{2}, and glide plane $c$ with glide vector (0,0,1/2) marked in the crystalline \emph{ac}-plane.\cite{Hahn89} (d) The Cr$^{5+}$ ($s=1/2$) spin dimer composed of two CrO$_4$ tetrahedra with local inversion center $-$\emph{1}. Green, red, and gray spheres denote Cr, O, and Sr/Ba ions, respectively.}
\end{figure}

A first possibility is to consider the small static spin-orbit
corrections to the Heisenberg coupling (spin anisotropies), the
largest of them in $s=1/2$ systems being the
Dzyaloshinskii-Moriya coupling.
Such an interaction was suggested to explain the origin of the
observed transitions in \BaCrO\,on the condition of a putative lower crystal
symmetry,\cite{Kofu09b} lower than the observed one.
The crystal structure of \SrCrO~and \BaCrO~is
hexagonal at room temperature with space group
\emph{R}$\bar{3}$\emph{m}.\cite{Cuno89,Buschbaum72} Each Cr$^{5+}$ ion
with a single 3\emph{d} electron is surrounded by an oxygen
tetrahedron. The Jahn-Teller distortion leads to a structural phase
transition to a low-temperature monoclinic structure with space group
$C2/c$, see
Fig.~\ref{Fig:STScheme}(a).\cite{Chapon08,Kofu09a,Wang12} Two adjacent
CrO$_4$ tetrahedra along the hexagonal $c_h$ direction form a spin dimer with an inversion center
[Fig.~\ref{Fig:STScheme}(a)(d)].\cite{Chapon08,Aczel08,Islam10}

Here we
investigate the singlet-triplet transitions both in \SrCrO~and \BaCrO~by measuring
high-field ESR transmission spectra with different radiation
polarizations and external magnetic field orientations. We argue that these transitions may
occur in the absence of an assumed static symmetry breaking,\cite{Kofu09b} provided that
the Dzyaloshinskii-Moriya interaction is \textit{dynamical},
\textit{i.e.}, that dynamical lattice distortions break the symmetry
instantaneously. In this case, the transitions would result from
exciting the electric dipoles formed by the ions of the
spin dimers by the electric field of the radiation. Such electric field induced transitions have appeared in different contexts, \emph{e.g.}, as an explanation of \textit{umklapp} $q \approx \pi$
transitions,\cite{Mitra} magnon-like forbidden transitions in
spin-gapped systems,\cite{Cepas01,Cepas04,Room04} or
electromagnon excitations in multiferroic compounds.\cite{Pimenov} In these latter works, the coupling to the phonons plays an important role and can directly contribute to spin relaxation process.\cite{Eremin08} An explicit proof of
the electric vs. magnetic dipole character is not always
possible, but was given in SrCu$_2$(BO$_3$)$_2$.\cite{Room04}


\section{Experimental details}
High-quality single crystals of \SrCrO~and \BaCrO~were grown by the floating-zone
method as described in Refs. \onlinecite{Islam10} and \onlinecite{Aczel08}, and characterized in detail.\cite{Castro10,Wang11,Castro12,Aczel07,Aczel09,Aczel09PRB,Wulferding11}
At the Dresden High Magnetic Field Laboratory, a tunable-frequency ESR spectrometer equipped with a 16~T superconducting magnet is employed (similar to that described in Ref.~\onlinecite{spectrometer}). Backward Wave Oscillators (BWOs) and VDI microwave sources (product of Virginia Diodes Inc.) were used as tunable sources of mm- and sub-mm wavelength radiation.
Polarized ESR experiments were performed in Voigt geometry with home-made grid polarizers glued directly on the sample. For the Voigt geometry the propagating vector of the electromagnetic wave is aligned perpendicular to the external magnetic field.
High-field transmission experiments in Augsburg were performed in Voigt geometry
with BWOs covering
frequencies from 115 GHz to 1.4 THz and a magneto-optical
cryostat (Oxford Instruments/Spectromag) with applied magnetic
fields up to 7 T.
Single crystals of \BaCrO~and \SrCrO~with typical sizes $4\times2\times0.2-1$ mm$^3$ were measured in the high-field ESR experiments.
The crystals are aligned with respect to the hexagonal axes $a_h$, $b_h$, and $c_h$. In the following, the monoclinic axes are noted as $a$, $b$ and $c$.

\section{Experimental Results and Discussion}

\subsection{Spin triplet excitations}
The spin dimers form a layered structure stacked in an \emph{ABAB} sequence in the low-temperature phase
[Fig.~\ref{Fig:STScheme}(a)(b)]. The intra-dimer
exchange interaction $J_0$ (5.5~meV in \SrCrO~and 2.4~meV in \BaCrO) is larger than the inter-dimer
interactions.\cite{Castro10,Kofu09a} Due to the double-layer structure, the triplet excitations have two branches, \emph{i.e.}, an acoustic mode $\omega^+$ and an optical mode $\omega^-$,
corresponding to the ``$q=0$'' and ``$q=\pi$'' phase difference between spin dimers of adjacent layers, respectively.\cite{Leuenberger84,Haley72} The excitation energies at the zone center are given by a random phase approximation calculation\cite{Kofu09a,Castro10}
\begin{equation}
\omega^{\pm}=\sqrt{J_0^2+J_0\gamma^{\pm}}
\end{equation}
where $\gamma^{\pm}=2(J'_{2}+J''_{2}+J'''_{2})\pm[(J'_{1}+J''_{1}+J'''_{1})+(J'_{4}+J''_{4}+J'''_{4})]$.
The different exchange constants $J_{i}$ denote intra- and inter-layer interactions as illustrated in
Fig.~\ref{Fig:STScheme}(b) and were extracted from inelastic neutron scattering.\cite{Kofu09a,Castro10} Both modes are triplets and split in an external magnetic field with energies $\omega^{\pm}(H) =\omega^{\pm} + g\mu_{B}HS^z$ with $S^z=0,\pm 1$ [Fig.~\ref{Fig:AcOpMode_SCO}(a)]. This picture is confirmed by unpolarized transmission ESR measurements up to 16~T for \SrCrO~and 13~T for \BaCrO~as shown in Fig.~\ref{Fig:AcOpMode_SCO}(b) and Fig.~\ref{Fig:AcOpMode_BCO}(a), respectively. Note that excitations to the $S^z=0$ levels are absent because of the sweeping field technique.  The obtained excitation energies extrapolated to zero field are $\omega^{+}=1.47$~THz ($6.1$~meV) and $\omega^-=1.24$~THz ($5.1$~meV) for \SrCrO, and $\omega^{+}=563$~GHz ($2.3$~meV) and $\omega^-=399$~GHz ($1.6$~meV) for \BaCrO~with $g=1.94(3)$ for both compounds. The energies are perfectly consistent with inelastic neutron experiments,\cite{Kofu09a,Castro10} which also measured their triplet nature.\cite{Kofu09a} The acoustic mode has higher energy than the optical mode in both compounds because the inter-dimer couplings are dominated by antiferromagnetic exchange interactions.\cite{Kofu09a,Castro10} In the extrapolation, there is no zero field splitting of the different $S^z$ components detectable within the experimental uncertainty.


\subsection{Polarization analysis}
\label{hf}
\begin{figure}[t]
\centering
\includegraphics[width=85mm,clip]{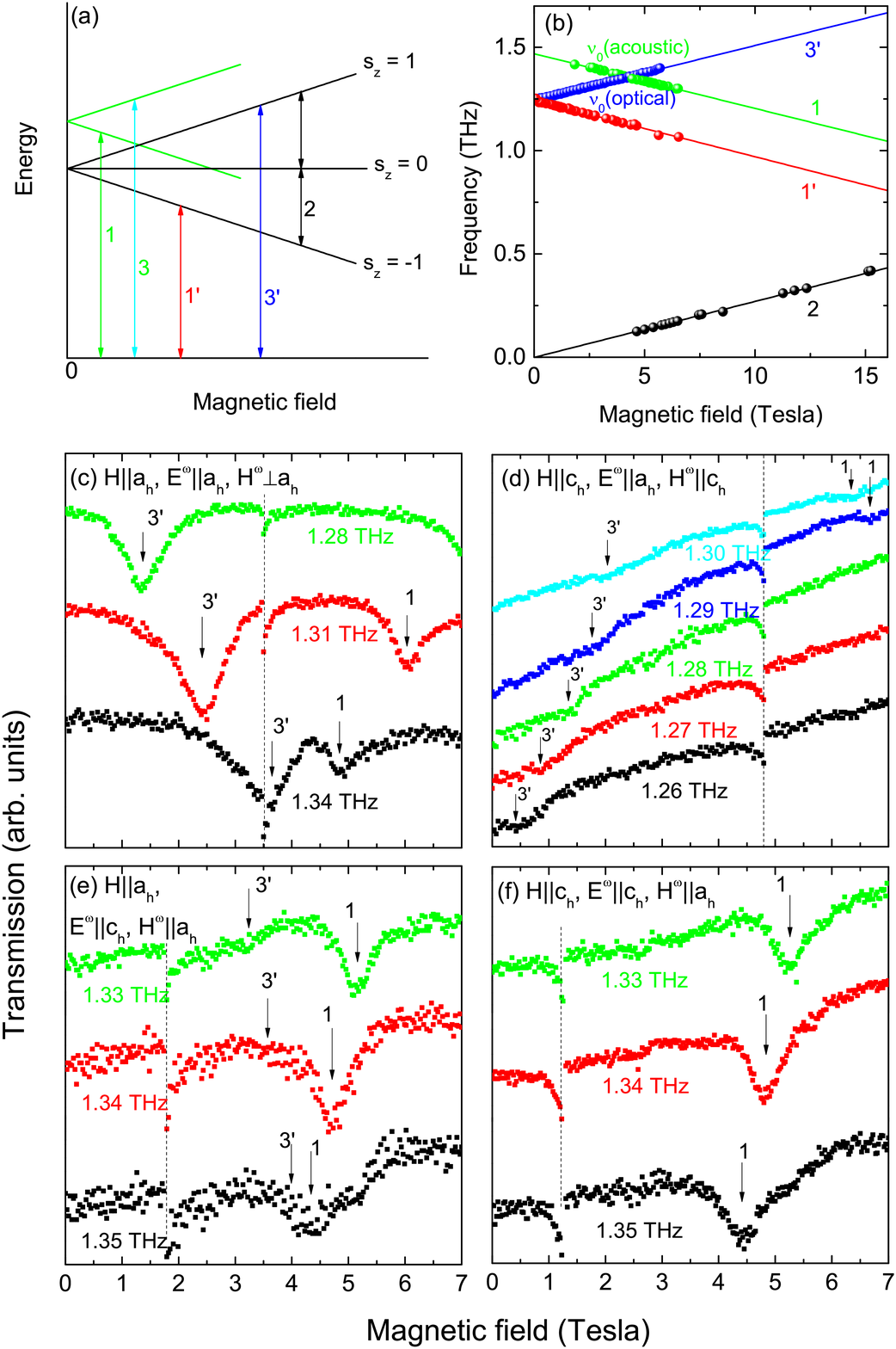}
\vspace{2mm} \caption[]{\label{Fig:AcOpMode_SCO} (Color
online) \SrCrO~ (a) Schematic singlet-triplet excitations in an external magnetic field. 1 and 3 are the acoustic modes, and 1' and 3' are the optical modes.
The two types of modes originate from the two inequivalent layers in the unit cell.
Mode 2 denotes the intra-triplet excitations.
The singlet-triplet excitations are measured with no polarization analysis (b) and with polarization analysis ($E^\omega, H^\omega$) (c)-(f)
for different orientations of applied magnetic field $H$ at 2~K.
The vertical dashed lines indicate artifacts due to spark lines from the BWOs.
}
\end{figure}

\begin{figure}[t]
\centering
\includegraphics[width=85mm,clip]{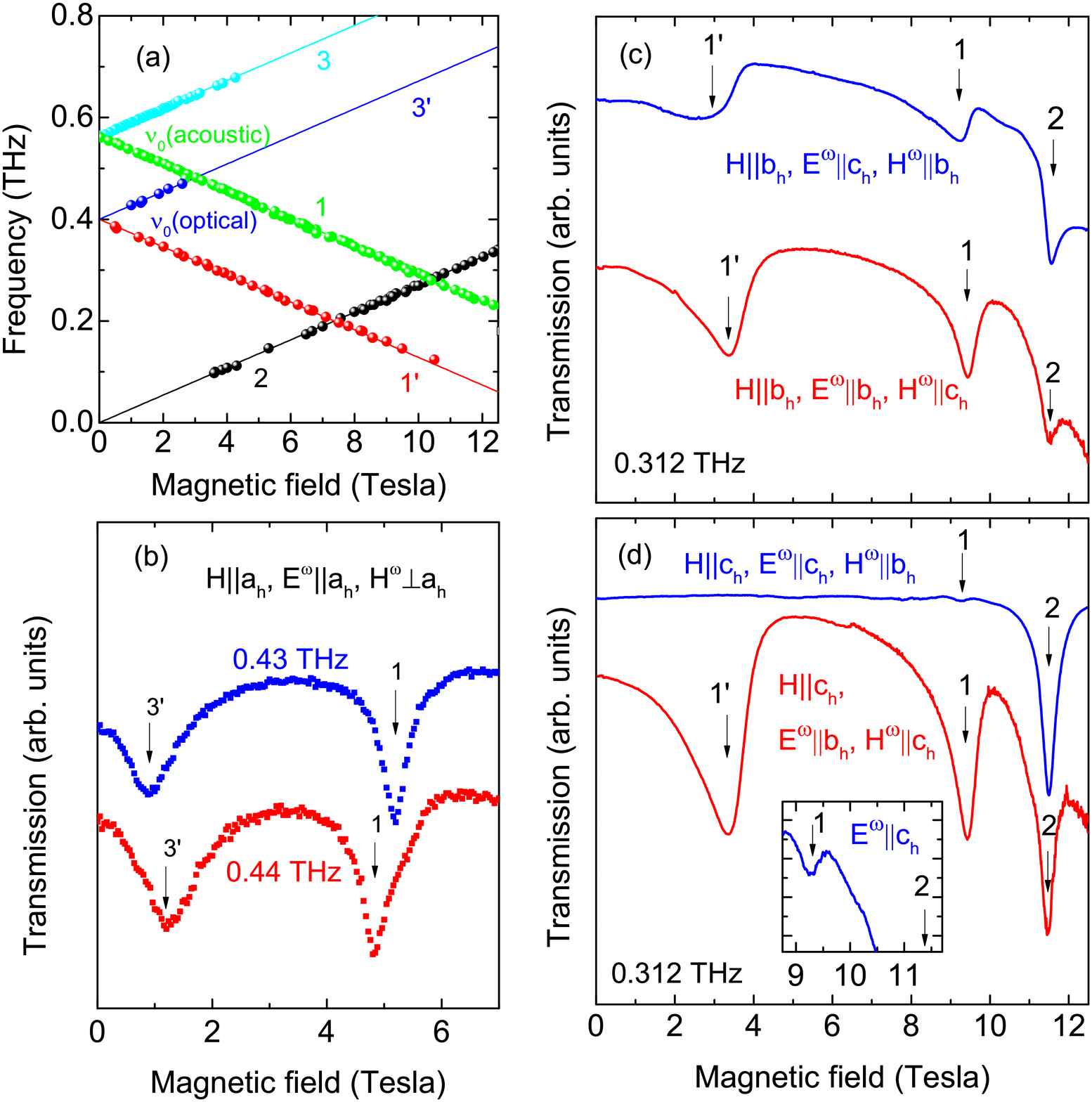}
\vspace{2mm} \caption[]{\label{Fig:AcOpMode_BCO} (Color
online) \BaCrO~(a) Singlet-triplet excitations. Polarized transmission spectra measured at 2~K (b) and at 1.4~K (c,d). In (c), $H^\omega$ is parallel to the $a_hb_h$-plane. Inset: enlarged view of the spectrum for $H \| c_h$, $E^\omega \| c_h$, and $H^\omega \| b_h$.}
\end{figure}

ESR transmission experiments were performed for different directions of the polarization of the electromagnetic radiation and orientations of the external magnetic field with respect to the hexagonal axes of \SrCrO~and \BaCrO~(Figs.~\ref{Fig:AcOpMode_SCO},\ref{Fig:AcOpMode_BCO}). The field-dependence of the singlet-triplet excitations has been determined in the unpolarized experiments [Fig.~\ref{Fig:AcOpMode_SCO}(b) and Fig.~\ref{Fig:AcOpMode_BCO}(a)]. One can accordingly identify the excitations in the spectra measured with polarized radiations.

Figures~\ref{Fig:AcOpMode_SCO}(c)-(f) show polarized ESR transmission spectra of \SrCrO~measured with respect to the hexagonal axes. The acoustic mode 1 and the optical mode 3' are observed and are in agreement with the unpolarized spectra shown in Fig.~\ref{Fig:AcOpMode_SCO}(b). In a finite external magnetic field, the acoustic modes and optical modes exhibit different polarization dependencies: the acoustic modes are observed for all the polarizations, while the optical modes are absent for $E^\omega \| c_h$, $H \| c_h$ [Fig.~\ref{Fig:AcOpMode_SCO}(f)].

Figure~\ref{Fig:AcOpMode_BCO}(b)-(d) shows the polarized ESR transmission spectra of \BaCrO~with various polarization configurations. One can identify the acoustic mode 1, the optical modes 1',3', and the intra-triplet mode 2 in accord with Fig.~\ref{Fig:AcOpMode_BCO}(a). The most prominent feature is also that the optical modes are absent only for $E^\omega \| c_h$, $H\| c_h$, while the acoustic modes are observed for all the polarizations in \BaCrO.

Considering the geometric relations $c_h=(a+3c)/2$,\cite{Note1} $b_h=(-a+b)/2$, $a_h=(a+b)/2$ between hexagonal and monoclinic axes,\cite{Castro10} the contribution of the monoclinic axes $a$ and $b$ are probed simultaneously in the polarization measurements for $E^\omega \| b_h$ and for $E^\omega \| a_h$. Thus, \SrCrO~and \BaCrO~exhibit the same polarization dependent selection rules, which are summarized in Table \ref{tab:Exp_selection_rule_BCO}.

\begin{table}[t]
\caption{\label{tab:Exp_selection_rule_BCO}  \emph{Experimentally} observed spin singlet-triplet excitations in the polarized ESR spectra in the \emph{hexagonal} notations. The acoustic mode $1$ and optical mode $1'$, $3'$ are noted as '$0$' and '$\pi$', respectively, for short.}
\begin{ruledtabular}\vspace{0.5mm}
\begin{tabular}{c@{\hspace{2em}}*{4}{c}}
$E^\omega \backslash H$   &$a_h$                                     &$b_h$                       &$c_h$
\vspace{0.5mm}\\\hline
\vspace{0.5mm}
$a_h$                     &$0+\pi$                                &                             &     $0+\pi$          \\
                          &[Fig.~\ref{Fig:AcOpMode_SCO}(c),~\ref{Fig:AcOpMode_BCO}(b)]       &                             &[Fig.~\ref{Fig:AcOpMode_SCO}(d)]
\vspace{0.5mm}\\\hline
\vspace{0.5mm}
$b_h$                     &                                          &$0+\pi$                   &$0+\pi$            \\
                          &                                          &[Fig.~\ref{Fig:AcOpMode_BCO}(c)]                            &[Fig.~\ref{Fig:AcOpMode_BCO}(d)]
\vspace{0.5mm}\\\hline
\vspace{0.5mm}
$c_h$                     &$0+\pi$                                         &$0+\pi$                  &$0$    \\
                          &[Fig.~\ref{Fig:AcOpMode_SCO}(e)]                                         &[Fig.~\ref{Fig:AcOpMode_BCO}(c)]    &[Fig.~\ref{Fig:AcOpMode_SCO}(f),~\ref{Fig:AcOpMode_BCO}(d)]
\end{tabular}
\end{ruledtabular}
\end{table}

\subsection{Discussion}

A magnetic dipole singlet-triplet excitation in a single dimer can be observed when there is an intra-dimer static Dzyaloshinskii-Moriya interaction. Since this is forbidden in \SrCrO~and \BaCrO~due to the inversion center in the middle of the bonds in the Cr$_2$O$_8$ dimers [Fig.~\ref{Fig:STScheme}(d)], the observed transitions have been ascribed to additional undetected lattice distortions that break the inversion symmetry.\cite{Kofu09b} However, other mechanisms may be at play which do not need to invoke additional distortions.

One can keep the magnetic dipole mechanism in a pure magnetic model and consider the weaker dimer-dimer interactions. Static inter-dimer Dzyaloshinskii-Moriya interactions along the superexchange paths of $J_1$, $J_2$, and $J_4$ (but not $J_3$ which has an inversion center) can mix the singlet and triplet states, and allow the magnetic dipole transitions.\cite{Sakai00}
However, we consider it more likely that an electric dipole mechanism involving the strongest exchange interactions dominates the excitation intensities.
This is because (1) a static singlet-triplet mixing generally bends the magnetization curves (this is not observed in the present compounds\cite{Aczel09,Kofu09b}) and opens a zero-field splitting (which is not detectable here and less than a few GHz); and (2) rough estimates of relative intensities tend to favor the electric dipole mechanism,\cite{Cepas04} \textit{a fortiori} when the magnetic dipole transitions are plagued by an additional small parameter, $\sim J'/J_0$.


We consider an electric-dipole coupling between radiation and spins,
\begin{equation}\label{Eq:Woperator}
W=\sum_{i,\alpha,\beta} E^\omega_\alpha A_{\alpha\beta} (\mathbf{S}_{i1}\times \mathbf{S}_{i2})_\beta
\end{equation}
where $i$ refers to the spin dimers, and $A_{\alpha\beta}$ is a coupling constant that couples the electric field of the radiation along the $\alpha$ direction with the vector product of spins along the $\beta$ direction, that has the Dzyaloshinskii-Moriya symmetry. This is the simplest spin operator for $s=1/2$ systems that breaks spin rotation symmetry (but not the time-reversal symmetry), and allows for a non-zero matrix element for singlet-triplet transitions.

\begin{figure}[t]
\centering
\includegraphics[width=34mm,clip]{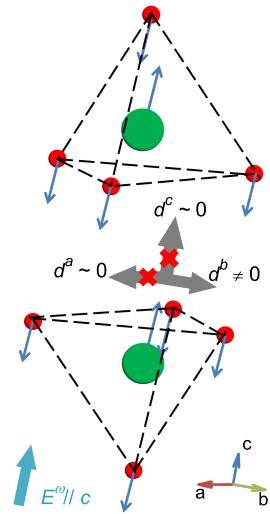}
\vspace{2mm} \caption[]{\label{Fig:DMvectors} (Color
online) Illustration of the coupling constants $A_{ca}\sim0$, $A_{cb}\neq0$, and $A_{cc}\sim0$.
The spin dimer Cr$_2$O$_8$ distorts in the radiation electric field $E^\omega \| c$ with associated instantaneous Dzyaloshinskii-Moriya vectors $d^{a}$, $d^{b}$, and $d^{c}$.}
\end{figure}

Operators of the same form have been used to explain double magnon excitations in several antiferromagnetic systems,\cite{Elliott69,Gruninger96} and similar processes in dimer systems,\cite{Cepas01,Cepas04} but the microscopic mechanism remains unclear. In the latter case, it is assumed that an \textit{optical} phonon  breaks the inversion symmetry within the dimer instantaneously. Since the electronic hopping is much faster than lattice vibrations, superexchange takes place in a dimer with no inversion center. The Dzyaloshinskii-Moriya interaction is therefore generated thanks to the spin-orbit coupling, and is linear in the ionic displacements at first order.
Typically $A_{\alpha\beta}= D^\alpha d^\beta$ where $d^\beta$ is the instantaneous Dzyaloshinskii-Moriya vector and $D^\alpha$ the electric dipole of the unit cell created by the virtual phonon (see an example in Fig.~\ref{Fig:DMvectors}).

\textit{Symmetry arguments}. The coupling constant $A_{\alpha \beta}$ must be constrained by the crystal symmetries of the lattice in the presence of the radiation electric field. In particular, they should be identical from unit cell to unit cell but may vary within the unit cell. The space group $C2/c$ has four symmetry elements, namely, identity \emph{1}, inversion center \emph{-1}, two-fold rotation axis \emph{2}, and \emph{ac}-glide plane \emph{c} with glide vector $(0,0,1/2)$, as shown in Fig.~\ref{Fig:STScheme}(c).\cite{Hahn89} The last two symmetry operations always transform the pseudo-vector $\mathbf{T}_{12} \equiv \mathbf{S}_{1}\times \mathbf{S}_{2}$ of the two dimerized spins in one layer onto one in the next layer, labeled with primes, as in Fig.~\ref{Fig:STScheme}(b). By using these symmetries, we constrain the coupling constants:
\begin{itemize}
\item $E^\omega \| a,c$: the \emph{ac} glide plane leaves $E^\omega$ invariant,
 transforms $T_{12}^b$ onto $T_{1''2''}^b$, and $T_{12}^{a,c}$ onto $-T_{1''2''}^{a,c}$ (Fig.~\ref{Fig:STScheme}). The coupling constants $A_{\alpha\beta}$ within the unit cell are therefore not independent but only the $\pm$ phases appear: the operator $W$ in Eq.~(\ref{Eq:Woperator}) will generate excitations to the acoustic and optical branches, respectively.
\item $E^\omega \| b$: the two-fold rotation axis is along the \emph{b} axis. It leaves $E^\omega$ invariant and transforms $T_{12}^b$ onto $-T_{1'2'}^b$, and $T_{12}^{a,c}$ onto $T_{1'2'}^{a,c}$: the operator $W$ will generate excitations to the optical and acoustic branches, respectively.
\end{itemize}
Based on the symmetries, this mechanism predicts the transitions to
occur at the energies of the acoustic and optical modes,
in perfect agreement with the energies observed for both
compounds. We emphasize that no static breaking of inversion symmetry is
needed.

A finite external magnetic field $\mathbf{H}$ splits the triplet modes
and each mode has its own intensity. If all the involved pseudo-vectors $\mathbf{T}_{ij}$ in the operator $W$ are parallel to the magnetic field $\mathbf{H} \| z$,
only the transition to the $S^z=0$ component of the triplet occurs, but it is invisible in the present high-field ESR setup. Therefore only the components of $\mathbf{T}_{ij}$ that are perpendicular to $\mathbf{H}$ generate observable transitions.
Taking into account the symmetry arguments given above, we infer the results given in Table~\ref{tab:partial_selection_rule_th}.

\begin{table}[h]
\caption{\label{tab:partial_selection_rule_th}  \emph{Theoretically} predicted excitations to the $S^z=\pm 1$ triplet components according to the dynamical mechanism in the \emph{monoclinic} notations. The acoustic mode and optical mode are noted here as '$0$' and '$\pi$', respectively, for short. The responsible matrix elements are given in the brackets following the corresponding modes. }
\begin{ruledtabular}\vspace{0.5mm}
\renewcommand\arraystretch{1.5}
\begin{tabular}{c@{\hspace{2em}}*{3}{c}}
$E^\omega \backslash H$   &$a$                                     &$b$                       &$c$
\vspace{0.2mm}\\\hline

\vspace{0.5mm}
$a$                     &$0(A_{ab}) +\pi(A_{ac})$                                    &$\pi(A_{aa},A_{ac})$                             &$0(A_{ab}) + \pi(A_{aa})$

\vspace{0.2mm}\\\hline
\vspace{0.5mm}
$b$                     &$0(A_{bc})+\pi(A_{bb})$                                        &$0(A_{ba},A_{bc})$                         &$0(A_{ba})+\pi(A_{bb})$

\vspace{0.2mm}\\\hline
\vspace{0.5mm}
$c$                     &$0(A_{cb})+\pi(A_{cc})$                                        &$\pi(A_{ca},A_{cc})$                  &$0(A_{cb})+\pi(A_{ca})$

\end{tabular}
\end{ruledtabular}
\end{table}

We cannot strictly test experimentally the selection rules given in Table~\ref{tab:partial_selection_rule_th}, because in the low-temperature phase the samples have three monoclinic twins rotated about the $c_h$ axis by an angle of 60$^o$ with each other.\cite{Castro10} As a consequence, a field applied along the hexagonal axis, $a_h$ or $b_h$, has components along \emph{a} and \emph{b}, thus mixing the selection rules. Furthermore, the hexagonal $c_h$-axis is tilted  from the \emph{c}-axis by 12$^o$, so that a field applied along $c_h$ has a main component along $c$ but also a small component along $a$. The mixing of the different components implies that both modes are predicted to have finite intensities in all the experimental configurations studied here, if the corresponding couplings $A_{\alpha \beta}$ (given in Table~\ref{tab:partial_selection_rule_th}) are non-zero.
However, the optical mode is not detected in the configuration $E^\omega \| c_h$, $H \| c_h$, see Table~\ref{tab:Exp_selection_rule_BCO}.
  If we ignore
  the difference between the $c_h$ and $c$ axes (which is small), the extinction of the optical mode for
  the configuration $E^\omega \| c_h$, $H \| c_h$ can be interpreted by postulating a
  vanishing (or weak) coupling constant $A_{ca}=0$, see
  Table~\ref{tab:partial_selection_rule_th}. Given
  that all the coupling elements are allowed by symmetry in the low
  temperature phase, it is difficult to prove that
  $A_{ca}$ vanishes.

We now discuss this extinction as the result of a
  possible approximate symmetry, originating from the higher symmetry
  of the high-temperature phase.
In the high-temperature phase, there are three mirror planes
at 120$^o$ bisecting the basal O-O bonds and containing the two Cr
ions of the dimer. Since $E^{\omega}$ along $c_h$ does not break these symmetries,
the instantaneous Dzyaloshinskii-Moriya vector should be perpendicular
to all, and hence would vanish. None of these mirror
  planes are exact symmetries in the low-temperature phase, but one of
  them coincides with the \emph{ac} plane and contains the two Cr
  ions.\cite{Hahn89,Chapon08} If we assume that this plane remains an
  approximate mirror plane, the
  instantaneous Dzyaloshinskii-Moriya vector would be along the $b$ axis, when $E^{\omega}\| a,c$ respects this
  symmetry, \emph{i.e.}, $A_{ab} \neq 0, A_{cb} \neq 0$. Therefore we would
  expect zero couplings $(A_{aa}=0,A_{ac}=0, A_{ca}=0,A_{cc}=0)$ if this symmetry were exactly
  preserved, or approximately zero if this symmetry were weakly
  broken. This gives a justification for the extinction of the
  optical mode in both compounds ($A_{ca} \approx 0$). It is also consistent with the weak
  intensity of the optical mode for the configuration $E^\omega \| c_h$, $H \| a_h$
  which is observed in \SrCrO~[Fig.~\ref{Fig:AcOpMode_SCO}(e)],
  provided that $A_{cc} \approx 0$ is weak.  However, in \BaCrO~with the configuration $E^\omega \| c_h$, $H \| b_h$
  [Fig.~\ref{Fig:AcOpMode_BCO}(c)]  the optical mode is much stronger. Within the present mechanism, this can be interpreted only with a finite $A_{cc} \neq 0$ (since
  $A_{ca} \approx 0$). This possibly implies that the symmetry discussed is more strongly broken in the low temperature phase in \BaCrO~than in \SrCrO.


Other mechanisms cannot be completely discarded by the present study, especially because it is difficult to assess their relative intensities. The present one, however, gives a consistent interpretation of the experimental results and would remain operative in systems with arbitrarily weak coupling between dimers with an inversion center.

\section{Conclusions}
By performing polarization-dependent high-field spectroscopy measurements at low temperature, we have found that the three-dimensional spin dimerized antiferromagnets \BaCrO~and \SrCrO~exhibit the same polarization dependence of spin singlet-triplet excitations, thus confirming the same spin symmetry in both compounds. Restricting to the isolated dimer picture, we have explored an electric dipole active mechanism that provides an explanation for the occurrence of both the acoustic and optical modes. In this mechanism, the previously assumed but symmetry-forbidden static Dzyaloshinskii-Moriya coupling is replaced by instantaneous couplings to the phonons. From the observed selection rules, some of these couplings are inferred to vanish (or to be weak), suggesting that the lattice symmetry that we have identified may in fact be only weakly broken across the structural phase transition.


\begin{acknowledgments}
We thank V. Tsurkan for help with sample preparation. We acknowledge partial support by the Deutsche Forschungsgemeinschaft via TRR 80
(Augsburg-Munich), Project DE 1762/2-1 and ZV 6/2-1, and also the support of the HLD at HZDR, member of the European Magnetic Field Laboratory.

\end{acknowledgments}

\end{document}